\documentstyle[prl,aps,multicol,epsf]{revtex}
\renewcommand{\narrowtext}{\begin{multicols}{2} \global\columnwidth20.5pc}
\renewcommand{\widetext}{\end{multicols} \global\columnwidth42.5pc} 

\begin{document}

\newcommand{\be}{\begin{equation}}
\newcommand{\ee}{\end{equation}}
\newcommand{\bea}{\begin{eqnarray}}
\newcommand{\eea}{\end{eqnarray}}
\newcommand{\nt}{\narrowtext}
\newcommand{\wt}{\widetext}

\title{Decoherence of one-dimensional flying qubits due to their cross-talk and imperfections}

\author{D. V. Khveshchenko}

\address{Department of Physics and Astronomy, University of North
Carolina, Chapel Hill, NC 27599}
\maketitle

\begin{abstract}
We study decoherence of propagating 
spin-$1/2$ excitations in generic (non-integrable
and/or disordered) spin chains. We find the relevant decoherence times to be shorter in
both the near-critical and diffusive regimes (if any), which fact 
might have important implications for the recently proposed 
spin chain-based implementations of quantum information processing.
\end{abstract}

\nt
In spite of its success in describing thermodynamics
and, more recently, spin/thermal transport,
the theory of quantum one-dimensional ($1D$) spin systems has been struggling to
develop a systematic approach to such non-equilibrium problems as, e.g.,
the time evolution of a certain (in general, non-stationary)
input state even in the simplest exactly solvable cases (see, e.g., Refs.\cite{Barouch}).
However, it is exactly this kind of problems that
have been brought about by the recent advents of spintronics 
and quantum information processing (QIP).

For example, many theoretical discussions of an abstract multi-qubit $(N\gg 1)$ register
invoke a generic spin-$1/2$ Hamiltonian \cite{QIP}
\be
H=\sum_{a=x,y,z}(\sum_{i=1}^N{B}^a_i{S}^a_i+\sum_{i,j=1}^NJ^a_{ij}{S}^a_i{S}^a_j)
\ee
while the usually proposed architectures have the topology of a $1D$, $2D$ or  
$3D$ array with the nearest-neighbor (NN) inter-qubit couplings (despite
a desireable reduction in the required number of qubits,
the higher connectivity networks may not be that easy to assemble in practice).

In the QIP context, the issue of paramount importance is that of quantum coherence 
which, if no precautions are taken,
can be easily destroyed by any coupling to an external dissipative environment. 
Moreover, even though in the absence of such a coupling the density 
matrix ${\hat \rho}(t)$ of the entire $N$-qubit system undergoes unitary evolution, 
any reduced density matrix (RDM) obtained by tracing out some of the degrees of freedom 
can still exhibit an ostensible decay caused solely by the qubits' "cross-talk".

In the QIP-related applications, one is primarily interested in the short-time
behavior which strongly diminishes any chances of the Poincare recurrences 
of the initial quantum state that might occur in the noiseless limit at longer times.
Furthermore, an arbitrarily weak coupling to a noisy environment eliminates 
any possible quantum states' revivals
and makes this "deterministic decoherence"
a truly irreversible behavior.

Contrary to the non-interacting case, however, the analysis
of decoherence in a system of many coupled qubits described by Eq.(1) can 
no longer be carried out in terms of the conventional
transverse/longitudinal relaxation times ($T_{1,2}$) of individual on-site spins. 
Instead, an adequate discussion must start out by identifying the proper basis of delocalized spin-$1/2$
states with definite momenta (hereafter referred to as "spinons") that span the Hilbert
space of the interacting multi-qubit system (1) and then proceed towards computing 
various $n$-spinon ($n=1,2,\dots$) RDMs given by the partial traces 
$Tr_{1,\dots,N-n}{\hat \rho}(t)$.

Besides providing an appropriate framework for the discussion of
decoherence, the spinon basis can also offer a natural implementation
of the notion of "flying qubits" proposed as a vehicle for performing such important
QIP tasks as teleportation or quantum key distribution \cite{QIP}.

To this end, in the present paper we evaluate the spinon decoherence rates 
in several situations that might be of interest for QIP and ascertain the prospects of
using spinons in the QIP protocols relying on the availability of flying qubits.

At first sight, the presence of unbound spinons in the spectrum of the Hamiltonian 
(1) may seem like a commonplace, as suggested by the possibility to readily cast Eq.(1)
in the form of the lattice Hamiltonian 
of spin-$1/2$ fermions with some short-ranged quartic interactions by 
applying a fermion representation of the spin operators 
(${\vec S}_i={1\over 2}\psi^\dagger_{i\sigma}{\vec \sigma}_{\sigma\sigma^\prime}
\psi_{i\sigma^\prime}$) complemented by a local single-occupancy constraint
($\psi^\dagger_{i\sigma}\psi_{i\sigma}=1$). However, the answer to the question as to whether or not
the "constituent" fermion operators 
$\psi_{i\sigma}$ correspond to any (possibly, approximate) 
eigenstates of the Hamiltonian (1) would strongly depend on the actual form of the latter.

The existence of deconfined spinons in
higher dimensions (or a lack thereof) had been the subject of a long ongoing debate 
which was rekindled by the discovery
of cuprates representing a new class of strongly correlated Mott insulators.
Despite an extensive work on the topic, however, only a handful of 
solid arguments in favor of the occurrence of
such excitations in a number of rather exotic models have been made so far
(see, e.g., \cite{UCSB} and references therein).

In contrast, many 1D spin systems do conform to the 
picture of a "spinon Fermi liquid" (SFL) whose properties appear to be
quite different from those of the spin-ordered states
described in terms of the conventional spin-$1$ magnons.
Incidentally, the interest in spin-chains as a promising 
layout of the practical quantum register has been bolstered 
by the reports \cite{QCP} of a massive buildup of quantum entanglement at 
those points in the parameter space of Eq.(1) 
where the system undergoes quantum phase transitions.

It should be noted, however, that, depending on the qubits' physical makeup,
the corresponding Eq.(1) may lack any particular spin symmetry, thus hampering  
the applicability of such customary 1D techniques 
as the Bethe ansatz, bosonization, and conformal field theory.
 
As an alternate approach that has been previously employed in 
various strongly correlated fermion systems, we adopt the 
semi-phenomenological spin-fermion model described by the action \cite{SFM}
$$
A=\sum_{k}
\int{dE\over (2\pi)}\psi^\dagger_{k\sigma}(E-\epsilon_k)\psi_{k\sigma}+
\sum_q\int{d\omega\over (2\pi)}S^a_q\chi^{-1}_{ab}(\omega,q)
$$
\be
S^b_{-q}+g\sum_{k,q}\int{dEd\omega\over (2\pi)^2}
\psi^\dagger_{k\sigma}{\hat \sigma}^a_{\sigma\sigma^\prime}{S}^a_q\psi_{k+q,\sigma^\prime}
\ee
written in terms of the continuum fermionic variables
which are no longer subject to the local single-occupancy constraint,
while their interactions are mediated by spin density fluctuations
governed by the dynamical spin susceptibility
$\chi^{ab}(\omega,{q})=\sum_{ij}\int dte^{i\omega t-iq|i-j|}<S^a_i(t)S^b_j(0)>$. 
 
The gapful fermion dispersion $\epsilon_k={\sqrt {v^2k^2+\Delta^2}}$ 
accounts for the main effect of the processes of backward scattering, 
and the gap $\Delta\ll J$ is assumed 
to vanish at certain values of the microscopic parameters in Eq.(1).  

Thus, the model (2) describes a gas of (for any $\Delta\neq 0$, massive) 
spinons that are still subject to a forward scattering, the singular part of which 
is controlled by the effective coupling constant $g$ 
measuring a deviation from the parent SFL state. In turn, the latter 
is thought to be composed of quasiparticles that
can be adiabatically related to the one-spinon states $|k\sigma>=\psi^\dagger_{k\sigma}|0>$
even in the presence of non-singular residual interactions that were omitted in Eq.(2).

Right at the quantum critical point (QCP) $\Delta=0$, and at 
all energies below a high-energy cutoff set by the average exchange coupling $J$
the $T=0$ susceptibility has a generic power-law form
\be
\chi^{ab}_{QCP}(\omega, {q})=
{\eta^{ab}(q)J^{2\alpha_{ab}}\over (v^2q^2-\omega^2)^{\alpha_{ab}}}
\ee
In the absence of spin-rotational invariance,
the exponents $\alpha_{ab}$ and the prefactors $\eta^{ab}(q)$
depend on the direction in spin space.
In the isotropic ($SU(2)-$) or axially ($U(1)$-) symmetric case, however, all (or some)
of the numerators vanish at zero transferred momentum $q$ 
in compliance with the conservation of the corresponding component of the total spin.

At temperatures above the gap 
($\Delta<T\lesssim J$) the susceptibility demonstrates a characteristic "quantum-critical" 
(QC) behavior 
\be
\chi^{ab}_{QC}(\omega,{q})=
{\chi^{ab}_0(T)\over {q^2\xi^2-(\omega-i\gamma_{QC})^2/\Omega^2}}
\ee
where $v/\xi\sim\Omega\sim\gamma_{QC}\propto T$ and
$\chi^{ab}_0(T)\propto (J/T)^{2\alpha_{ab}}$ \cite{Sachdev1}.

Notably, the temperature-dependent relaxation rate 
$\gamma_{QC}={1\over 2}Im {d\over d\omega}\ln\chi(\omega,0)|_{\omega\to 0}$ 
manifested by Eq.(4) can be present even in 
the host SFL state (i.e., for $g=0$).

At low temperatures ($T\lesssim\Delta$) the
off-critical susceptibility may also differ between the
"quantum disordered" (QD) and "renormalized classical" (RC) regimes \cite{Sachdev1}.
In either case, however, it can be cast in the form of 
a sum over the intermediate ($1,2,\dots$-fermion) states, 
the coherent (one-particle) component $\chi^{ab}_{coh}(\omega, {q})$
assumes a form similar to $\chi^{ab}_{QC}(\omega,{q})$
where $\chi^{ab}_0(\Delta)$ is now given by the same expression as in Eq.(4)
with the temperature substituted by $\Delta$, while the relaxation rate
$\gamma_{QD,RC}\propto Te^{-\Delta/T}$ \cite{Sachdev1,Jolicouer}
is suppressed as compared to its QC counterpart. 

In contrast to the pole-like structure of $\chi^{ab}_{coh}(\omega, {q})$, a generic 
incoherent two-particle contribution 
\be
\chi^{ab}_{inc}(\omega,{q})=
{\eta^{ab}(q)\over {(v^2q^2-\omega^2)^{\beta_{ab}}
(v^2q^2+4\Delta^2-\omega^2)^{1-\beta_{ab}}}}
\ee 
features either a singularity ($\beta_{ab}<1$) or 
a cusp ($\beta_{ab}>1$) at the $\omega=2\Delta$ threshold.

In what follows, we consider a representative example of the spin-isotropic system 
supporting massive spinons that is provided by the frustrated Heisenberg model (${J}^a_{ij}=
J\delta_{j,i\pm 1}+J^\prime\delta_{i,j\pm 2},~~{B}^a_i=0$).  
 
In this model (which can also be mapped onto a two-chain zigzag ladder), 
a sufficiently strong ($J^\prime>J^\prime_c\approx 0.24 J$)
next-nearest-neighbor (NNN) exchange gives rise to a spontaneous 
dimerization and opens up a gap $\Delta\propto J\exp(-const.J/J^\prime-J^\prime_c)$.

For all $J^\prime\leq J^\prime_c$ the system remains in the same critical state as  
the standard NN Heisenberg model described by the $SU(2)_{k=1}$ Wess=Zumino-Witten theory. 
Accordingly, the susceptibility given by the hydrodynamical expression 
$\chi^{ab}_{coh}(\omega,{q})=\delta^{ab}{v^2q^2/(v^2q^2-\omega^2})$ 
shows neither a gap, nor an anomalous exponent.

The effect of the NNN and other frustrating exchange interactions 
which render spinons massive can also be studied in the framework of the sine-Gordon (SG)
model where the spinons can be identified with the SG (anti)solitons \cite{Essler1}.

The incipient SFL instability with respect to dimerization occurs when the (anti)solitons 
become degenerate with some of their bound states ("breathers"), 
thus resulting in the formation of massive triplets (prototypes of "flying qutrits").

In the gapped regime, the low-$T$ susceptibility takes the form (5)
where $\eta_{ab}(q)\propto q^2$ and $\beta_{ab}={3\over 2}\delta_{ab}$, 
thereby indicating that the forward scattering remains non-singular 
and the (anti)solitons largely preserve their integrity.
This expectation is supported by our calculation of the spinon 
decoherence rates (see Eq.(12) below). 

Nonetheless, by analogy with the stability criteria 
of the conventional Fermi liquid,
one might expect that a singular (long-ranged and/or retarded) spin fluctuation-mediated 
interaction could make the spinons ill-defined and give rise to a 
critical "spinon non-Fermi liquid" state characterized by anomalous exponents.

The subsequent analysis shows that such a behavior can indeed 
occur in spin-anisotropic systems, a popular exactly solvable example of which 
is presented by the $XY$-model in transverse field
(${J}^{x}_{ij}=-J{1\over 2}(1+\delta)\delta_{j,i\pm 1}, ~ 
{J}^{y}_{ij}=-J{1\over 2}(1-\delta)\delta_{j,i\pm 1}, ~ {B}^z_i=B$). 

After the Jordan-Wigner transformation 
followed by the Bogoliubov rotation, the $XY$-model with $\delta\neq 0$
gets transformed into a free gas of Majorana fermions 
with the velocity $v=J\delta$ and the gap $\Delta=B-J$ 
that vanishes at the Ising QCP \cite{Sachdev1}.

These excitations possess no conserved physical spin
and can be thought of as domain walls ($Z_2$-kinks) between two locally 
degenerate ground states. The fact that the kinks 
lack a simple representation in terms of the constituent fermions $\psi_{k\sigma}$
implies that the corresponding one-spinon RDM 
${\hat \rho}(x-y,t)=<\psi^\dagger_\sigma(x,t){\psi}_{\sigma^\prime}(y,t)>$
should exhibit strong decoherence (see Eq.(13) below). 

In support of this expectation, the critical susceptibility of the model in question 
is now given by Eq.(3) where $\eta^{ab}(0)\neq 0$ and
the anomalous exponents depend on $\delta$ 
(e.g., $\alpha_{xx}=3/4$ and $7/8$ for $\delta=0$ and $1$, respectively). 

In the QD regime ($B>J$), the functions $\chi^{xx,yy,xy}$ feature
a pole-like structure, while in the RC domain ($B<J$) they exhibit a kinematic 
square-root singularity ($\beta_{xx, yy, xy}=1/2$) at the $2\Delta$ threshold.
In contrast, $\chi^{zz}$ displays a cusp ($\beta_{zz}=3/2$)
and remains incoherent for all values of
the transverse field \cite{Essler2}.

The unfying framework of the spin-fermion model (2) 
also allows one to investigate the effects of such qubits' imperfections as inhomogeneous local 
field or exchange couplings that vary from one site to another.
In the presence of a static disorder, the $SU(2)$-invariant low-energy 
($\omega, vq\lesssim\gamma_{tr}$) susceptibility acquires a diffusive pole
\be
\chi^{ab}(\omega, {q})={\chi^{ab}_uD_sq^2\over {D_sq^2+i\omega}}
\ee   
where the spin diffusion coefficient $D_s=v^2/2\gamma_{tr}$ is determined by the 
disorder-induced transport rate $\gamma_{tr}$. It is worth mentioning that
in a generic (non-integrable) spin chain $D_s$ is finite at $T>0$ 
even in the absence of disorder, consistent with a non-vanishing spin (thermal) resistivity
of such systems \cite{Sachdev2}.

Similar to the critical spin fluctuations,
the soft diffusive mode can prodive for a possible source of singular forward scattering 
between the spinons (see Eq.(14) below).

In either case, the effect of the low-energy 
spin-fluctuation exchange on the spinon evolution operator
${\hat U}^R(x,t)$ can be studied in the framework of
the continuum equation
\be
[i\partial_t-\epsilon(i\partial_x)+
i{\hat \sigma}^aS^a(x,t)]{\hat U}^R(x,t|{\vec S})={\hat {\bf 1}} \delta(x)\delta(t)
\ee
where ${\vec S}(x,t)$ represents an instantaneous configuration of the effective spin environment
that the propagating spinon is subjected to. 

The formal solution to Eq.(7) can be cast in the form
\be
{\hat U}(x,t|{\vec S})=
\int^\infty_0d\tau\sum_k\int{dE\over 2\pi}
e^{ikx-iEt-\tau(E-\epsilon_k)}
\ee
$$
P\exp[i\int^\tau_0e^{i\tau^\prime(\omega-\epsilon_{k+q}+\epsilon_q)}
d\tau^\prime\sum_q\int{d\omega\over 2\pi}
{\hat \sigma}^a{S}^a(\omega,q)(e^{i{q}{x}}-1)]
$$
where the matrix ordering $P$ is to be performed according to the time order
in which the $\hat {\vec \sigma}$-matrices appear
in the series expansion of the exponential operator.

Using Eq.(8) one can construct a one-spinon RDM 
given by the usual double-time Keldysh path integral
${\hat \rho}(x-y, t|{\vec S})=\int dudv
{\hat U}^R(x-u, t|{\vec S}){\hat \rho}^{(0)}
(u-v){\hat U}^R(v-y, -t|{\vec S})$
for a given realization of the spin environment and an initial density matrix
${\hat \rho}^{(0)}(x)$.

Upon statistically averaging ${\hat \rho}(x-y, t|{\vec S})$ over ${\vec S}(x,t)$,
we arrive at the expression
\be
{\hat \rho}(x, t)=\int dy{\hat L}(x-y,t){\hat \rho}^{(0)}(y)e^{-\Phi(x-y,t)}
\ee
where we singled out the (super)operator of unitary evolution
(hereafter $z=x-y+i0$)
$$
{\hat L}(z,t)={\hat {\bf 1}}\sum_ke^{ikz}\cos^2\epsilon_kt=
$$
\be
\Delta Im\sum_{\pm}{\sqrt {z\pm vt\over z\mp vt}}
K_1(\Delta{\sqrt {z^2-v^2t^2}})
\ee
At $\Delta=0$ the kernel ${\hat L}(z,t)={\hat {\bf 1}}\sum_{\pm}\delta(z\pm vt)$ 
describes a mere ballistic spreading
of the entanglement
present in the initial state 
(a "non-relativistic" counterpart of this 
behavior was studied in the case of the ferromagnetic
Heisenberg model in Ref.\cite{Eberly}).

Apart from the propagation, entanglement undergoes an apparent decay
characterized by the decoherence factor
$$
\Phi(z,t)={1\over 2}\sum_{q}\int{d\omega\over 2\pi}
{1-\cos(\omega-\epsilon_{k_*}+\epsilon_{k_*+q})t\over 
(\omega-\epsilon_{k_*}+\epsilon_{k_*+q})^2}
$$
\be
Im\chi^{aa}(\omega,q)(\coth{\omega\over 2T}-\tanh{\omega-\epsilon_k\over 2T})(1-\cos qz)
\ee
where $k_*\approx 1/z$, and the spinon Fermi distribution function eliminates
contributions of the bosonic modes
with energies $\omega\gtrsim T$ which, otherwise, would have resulted in a
residual decoherence at $T=0$.

Notably, Eqs.(9-11) bear a certain resemblance to the results of the 
semiclassical calculation of Refs.\cite{Sachdev2}
where the one-particle RDMs of the exactly solvable 
transverse field Ising model and $O(3)$ non-linear $\sigma$-model
were shown to factorize into a product of a quantum propagator 
and a relaxation factor, the latter becoming trivial at $T=0$.

Now, estimating Eq.(11) on the semiclassical trajectory $x-y=vt$, 
we can deduce a characteristic 
decoherence rate $\Gamma$ from the equation $\Gamma={d\over dt}\Phi(vt,t)|_{t=1/\Gamma}$. 

In the spin-isotropic case, we obtain
\bea
\Gamma_{sym}\propto g^2T, ~~~~ T>\Delta\nonumber\\
~~~~~~~ \propto g^2{T^2\over \Delta}e^{-\Delta/T}, ~~~~ T<\Delta
\eea
which is comparable to or smaller than the SFL relaxation rates 
$\gamma_{QC,QD,RC}$ caused by a non-singular scattering neglected in Eq.(2),
thus suggesting that the spinons remain well-defined elementary excitations. 

By contrast, in the totally anisotropic case 
the decoherence rate is strongly enhanced as compared to Eq.(12)
\bea
\Gamma_{asym}\propto g^2J^{2\alpha}T^{1-2\alpha} ~~~~ T>\Delta\nonumber\\
~~~~~~~ \propto g^2{J^{2\alpha}T^2\over \Delta^{1+2\alpha}}e^{-\Delta/T}, ~~~~ T<\Delta
\eea
The decoherence rates given by Eq.(13) become greater than
the SFL ones at $max[T,\Delta]<T_*\propto g^{1/\alpha}J$
which signals a rapid decay of any spinon-like wave packet with a definite spin $S=1/2$, 
thus indicating the difficulty of creating robust flying qubits
with the spin-anisotropic Hamiltonians. 

At temperatures below $T_*$ the analysis of decoherence requires one to proceed beyond 
the lowest order of the cluster expansion 
employed in the derivation of Eqs.(9-11). 

Besides, Eqs.(12,13) demonstrate that the decoherence exhibited by the one-spinon RDM 
can be suppressed by the spectral gap and, conversely, 
it tends to become stronger as the system is tuned towards a QCP. 
This conclusion complements the earlier analysis of the decohering 
effect of an external dissipative environment
which, unlike in the present case, remains unchanged \cite{DVK}. 

The decoherence rate also appears to be enhanced in the disordered case
where the susceptibility takes the form (6)
\be
\Gamma_{dif}\propto g^{4/3}T^{2/3}\gamma_{tr}^{1/3}
\ee 
The diffusive contribution (14) dominates over its ballistic counterpart (12) at
all temperatures up to $T_{**}\propto\gamma_{tr}/g^2$ where both terms 
are of order $\gamma_{tr}$, while 
below $T_{***}\propto g^4\gamma_{tr}$ the rate $\Gamma_{dif}$ becomes of 
order $T$ and a systematic account of higher orders of the cluster expansion is again
required. 

Besides being manifested by the single-spinon RDM, the rate (14) 
controls various amplitudes that are sensitive to the spinon phases such as, e.g., 
the "spin-Cooperon" amplitude associated with the pairs
of time-reversed trajectories that return to the starting point. Thus, $\Gamma_{dif}$ 
can also be extracted from the weak-localization correction 
to the low-temperature spin (thermal) conductivity
\be
\delta\sigma_s\sim\int^\infty_{\tau_{tr}}({D_s\over t})^{1/2}
e^{-\Phi(vt,t)}dt\propto (D_s\Gamma_{dif})^{1/2}
\ee
that should be contrasted with the lowest order (Drude) 
result $\sigma^{(0)}_s\propto D_s\chi_u$.

It was argued in Refs.\cite{Santos} that in the QIP context
the qubits' localization by disorder may prove advantageous, as 
far as the goal of creating well-localized spin-$1/2$ states is concerned. 
However, our results show that increasing the amount of disorder
would come at a price of greater decoherence rates.
 
To summarize, in the present work we studied the apparent spin decoherence
that occurs at $T>0$ in even a closed non-integrable 1D spin system.
This short-time "deterministic decoherence" can be 
revealed by, e.g., the RDMs of a finite number of spinons 
propagating in a dissipative environment
created by the rest of the interacting qubits.

We focused on the near-critical and diffusive regimes where the 
emergence of soft collective modes
(critical and diffusive, respectively) is more likely to cause a
singular forward spinon scattering and an ensuing rapid decay of the spinon RDMs.

In the case of a spin-isotropic Eq.(1) and no disorder, we found that spinon 
remains as good of a coherent elementary excitation as it is in the parent
SFL model.
By contrast, in the spin-anisotropic case and/or in the
presence of spin diffusion the decoherence rates are strongly enhanced,
consistent with the fact that the system does not support 
coherent spin-$1/2$ states with definite momenta.
We also observed that the decoherence
rates can be reduced by tuning the system away from an incipient
criticality and/or localization.

Lastly, our findings suggest that the enhanced decoherence
(as quantified by the spinon RDMs) in the near-critical and diffusive regimes
might significantly reduce any possible benefits of 
both the emergent massive entanglement \cite{QCP} and the disorder-induced localization of 
the flying qubits' \cite{Santos} that have been recently discussed
as potentially attractive features of the spin chain-based QIP designs.

This research was supported by ARO under Contract DAAD19-02-1-0049
and by NSF under Grant DMR-0071362.

\wt
\end{document}